\documentstyle[12pt,aaspp]{article}
\begin{document}

\title{MAPS OF THE MOLECULAR EMISSION AROUND 18 EVOLVED STARS}
\author{K. Z. Stanek\altaffilmark{1} and G. R. Knapp}
\affil{Princeton University Observatory, Princeton, NJ 08544-1001}
\affil{\tt e-mail: stanek@astro.princeton.edu, gk@astro.princeton.edu}
\altaffiltext{1}{On leave from N. Copernicus Astronomical Center,
Bartycka 18, Warszawa 00-716, Poland}
\author{K. Young and T. G. Phillips}
\affil{CSO, California Institute of Technology, Pasadena, CA 91125}
\affil{\tt e-mail: rtm@tacos.caltech.edu, phillips@tacos.caltech.edu}

\begin{abstract}

We present maps at $20''$ resolution of the molecular emission around
18 evolved stars (14~asymptotic giant branch stars, one supergiant,
two proto-planetary nebulae and one planetary nebula), mostly in the
$^{12}$CO(3-2) line. Almost all molecular envelopes appear to be at least
marginally resolved at this resolution. A substantial fraction of the molecular
envelopes show clear deviations from spherical symmetry in the form of
elliptical or bipolar envelopes. This indicates that there is a need to
implement non-spherical mass loss in current scenarios of the late stages of
stellar evolution, in particular on the asymptotic giant branch.

\end{abstract}


\section{Introduction}

     Stars in the latest stages of evolution, i.e. asymptotic giant branch
(AGB) and red supergiant stars, lose mass in the form of cool dusty
molecular winds at speeds of typically 5 to 40 $  km ~ s^{-1}$.  The
amounts of mass lost (up to $10^{-4} ~ M_{\odot} ~ yr^{-1}$) and the
duration of the mass loss phase (about $10^5$ years) are large enough
to ensure that this process dominates the evolution of the star at this stage.
Molecular line emission and dust continuum emission can both be observed
from these envelopes, and the use of these probes to measure mass loss has
become a well-developed and active field in the past fifteen years.
Work has been  directed towards such questions as measuring the rate of mass
loss and its effect on the evolution of the star, the gas to dust ratio, the
composition of the envelope (including the effects of photochemistry) and the
rate of mass return to the interstellar medium.  The models used to
analyze the observations usually contain the assumption that mass is lost
isotropically at a constant rate and at a constant outflow speed.
Observations with the new interferometers and submillimeter telescopes
have shown, however, that these assumptions are too simplistic.  Highly
evolved envelopes, such as that surrounding CRL618, show pronounced bipolar
structure and sometimes have a second very high-velocity wind close to the
star (Cernicharo et al.~1989; Gammie et al.~1989). Observations of CRL 2688
(Young et al.~1992) show the presence of two fast winds, one with an
outflow speed of $45\;km~s^{-1}$ and a second with a speed of at least
$100\;km~s^{-1}$. It is unlikely that these fast winds are driven by
radiation pressure -- the momentum in the starlight is insufficient.

Here we present observations showing that the assumptions
involving isotropy of the mass loss are also simplified. For 18
stars at late evolutionary stages we mapped the molecular envelopes,
mostly in the CO(3-2) line. Examining the CO(3-2) line, rather than a lower
$J$\/ transition, offers several advantages. The first is that higher $J$\/
 transitions have higher critical densities for collisional excitation.
For this reason emission from molecular clouds is less extensive
and the observations are less apt to be contaminated by Galactic emission.
Second, the antenna temperature of the CO(3-2) line will in general
be higher than in the lower $J$\/ lines. While the rotational transitions
above 3-2 may have even higher brightness temperatures, the Earths
atmosphere is much less transparent at high frequencies.

\section{Observations}

The data were obtained in January 1990 (R Scl, o Cet, U Cam, VY CMa, OH 231.8,
RS Cnc, R Leo, IRC +10216),
April--May 1990 (R Leo, V Hya, RT Vir, W Hya, R Cyg, CRL 2688, R Cas),
June and August 1992 (M 57)
and September 1993 ($\chi$  Cyg, V Cyg, IRC +40540, U Cam),
using the 10.4-m Caltech Submillimeter Observatory (CSO) telescope at
Mauna Kea (Hawaii, USA). We mostly observed the $^{12}$CO(3-2) line at
$\nu_o=345.796$ GHz, with a FWHM beam of  $20''$ and a main-beam
efficiency of 60\%.
The receiver was a double sideband SIS detector with a temperature of about
200 K and the zenith optical depth was 0.1 to 0.2, giving effective
single-sideband system temperatures of about 700 K
for observations made at elevations above $  50$ degrees.
The spectral line backend was a 1024 channel acousto-optical
spectrograph (AOS) of bandwidth 500 MHz, giving a channel spacing of 0.43
$  km~s^{-1}$ and a velocity resolution of 0.8 $  km~s^{-1}$.
The receiver detects both sidebands, and in some of the observations
the emission from other spectral lines is detectable and strong enough
to allow mapping also in these lines.  The CS(7--6)
line at $\nu_o$ = 342.883 GHz is detected in the carbon stars
IRC 10216, CRL 2688 and V Cyg. The $\rm ^{29}SiO$(8-7) line
at $\nu_o$ = 342.9791 GHz is seen from the oxygen stars R Leo and W Hya.

The individual observations were calibrated and those with bad
(strongly curved) baselines or high r.m.s. noise levels were
rejected.  The individual observations were then averaged, weighted
appropriately by the r.m.s. noise, and a first or second order polynomial
baseline was removed from the spectrum by
fitting it to regions of the spectrum judged to be free of line emission.
The data for the first and last fifty or so channels of the 1024 channel
AOS used at CSO were noisy and were not used in the fit.

Pointing was found to be correct within $5''$, but for some stars
uncertainties in the catalog position resulted in the shift of the
observed position of maximum flux. For maps which are the result of
averaging over several individual maps, the telescope pointing was checked
before each map was made.

\section{Results}

Some characteristics of the mapped stars are presented in Table~1.
We observed 14 AGB stars, one supergiant star (SG),
two protoplaneteary nebulae (PPNe) and one planetary nebula (PN).
Spectral and variability types  were obtained using the SIMBAD database.
Coordinates and chemical types were taken from the overview
of CO observations by Loup et al.~(1993).

\begin{planotable}{lllllllr}
\tablewidth{39pc}
\tablecaption{Observed Stars}
\tablehead{
\colhead{Name}	& \colhead{Other}	&
\multicolumn{2}{c}{Coordinates (1950)}  &
\colhead{Spe.}	& \colhead{Class}  	&
\colhead{Chem.}	\\
\colhead{}	& \colhead{name}	&
\colhead{$\alpha$ (h m s)} & \colhead{$\delta$ (\deg\  ' ")} &
\colhead{type} 	& \colhead{}  		&
\colhead{type}	}

\startdata
R Scl		& IRC $-$30015	& 01 24 40.02	& $-32\;48\;06.8$	&
C6II		& SRa		& C		\nl
o Cet	        & Mira		& 02 16 49.11	& $-03\;12\;22.4$	&
M7IIIe		& Mira		& O		\nl
U Cam		& IRC +60124	& 03 37 29.09	& $+62\;29\;18.8$	&
C6		& SRb		& C		\nl
VY CMa		& IRC $-$30087	& 07 20 54.74	& $-25\;40\;12.3$	&
M3-4II		& SG(Lc)	& O		\nl
OH 231.8+4.2	& QX Pup	& 07 39 58.90	& $-14\;35\;44.0$	&
M6I-M9III	& PPN		& O		\nl
RS Cnc		& IRC +30209	& 09 07 37.80	& $+31\;10\;05.0$	&
M6IIIase	& SRc		& O		\nl
R Leo  		& IRC +10215	& 09 44 52.24 	& $+11\;39\;40.4$	&
M8IIIe		& Mira		& O		\nl
IRC +10216	& CW Leo	& 09 45 14.89	& $+13\;30\;40.8$	&
Ce		& Mira		& C		\nl
V Hya		& IRC $-$20218	& 10 51 37.27	& $-21\;15\;01.1$	&
C9I		& SRa		& C		\nl
RT Vir		& IRC +10262	& 13 00 05.87	& $+05\;27\;15.0$	&
M8III		& SRb		& O		\nl
W Hya		& IRC $-$30207	& 13 46 12.08	& $-28\;07\;08.8$	&
M8e		& SRa		& O		\nl
M57		& Ring Nebula	& 18 51 43.70	& $+32\;57\;56.2$	&
?p		& PN		& \nodata	\nl
R Cyg		& IRC +50301	& 19 35 28.69	& $+50\;05\;11.7$	&
S6e		& Mira		& S		\nl
$\chi$  Cyg	& IRC +30395	& 19 48 38.53	& $+32\;47\;09.9$	&
S7e		& Mira		& S		\nl
V Cyg		& IRC +50338	& 20 39 41.42	& $+47\;57\;43.2$	&
C6e		& Mira		& C		\nl
CRL 2688	& Egg Nebula	& 21 00 19.90	& $+36\;29\;45.0$	&
F5Iae		& PPN		& C		\nl
IRC +40540	& LP And	& 23 32 01.30	& $+43\;16\;27.0$	&
C		& \nodata	& C		\nl
R Cas		& IRC +50484	& 23 55 52.07	& $+51\;06\;37.3$	&
M7IIIe		& Mira		& O
\end{planotable}

{\small
\begin{planotable}{lrrrrrrrr}
\tablewidth{40pc}
\tablecaption{Observational Results}
\tablehead{
\colhead{Name}		& \colhead{$T_{max}$} &
\colhead{Flux$^{\rm A}$}	& \colhead{$V^{\rm B}_{lsr}$}	&
\colhead{$V^{\rm B}_{e}$}	& \colhead{$\gamma^{\rm B}$}	&
\colhead{$V_{wing}$} 	& \colhead{$\sigma^{\rm C}_1$}
& \colhead{$\sigma^{\rm C}_2$} \\[.2ex]
\colhead{} 			& \colhead{[$\,K\,$]}		&
\colhead{[$\,K\;km\;s^{-1}\,$]}  	& \colhead{[$\,km\;s^{-1}\,$]} 	&
\colhead{[$\,km\;s^{-1}\,$]} 	& \colhead{}  			&
\colhead{[$\,km\;s^{-1}\,$]} 	& \colhead{[$\,K\,$]} & \colhead{[$\,K\,$]}}
\startdata
R Scl	  & 2.4	    & 58.7    & --19.0    &
15.2      & 0.1     & 19.0    & 0.08	  & 0.27      \nl
o Cet	  & 15.2    & 88.1    & 46.6      &
7.0$^{\rm D}$   & $18^{\rm D}$  & 10.0    & 0.06	  & 0.19      \nl
U Cam	  & 1.5	    & 45.6    & 4.9       &
24.5      & 0.4	    & 29.0    &	0.04	  & 0.22      \nl
VY CMa	  & 2.6	    & 159.6   & 17.6	  &
32.6	  & 0.4     & 92.0    &	0.06	  & 0.25      \nl
OH 231.8+4.2  & 1.9	    & 140.8   & 32.0$^{\rm D}$  &
65.0$^{\rm D}$  & $11^{\rm D}$  & 125.0   &	0.07	  & 0.19      \nl
RS Cnc	  & 4.3	    & 29.8    & 6.8       &
4.8$^{\rm D}$   & $4.6^{\rm D}$ & 10.2    &	0.07	  & 0.16      \nl
R Leo  	  & 2.8	    & 31.1    & --0.9     &
7.9       & 2.1	    & \nodata &	0.05	  & 0.15      \nl
R Leo$^{\rm E}$ & 0.6	    & 3.8     & \nodata   &
6.0$^{\rm F}$   & \nodata & \nodata &	0.05 	  & 0.15      \nl
IRC+10216 & 42.0    & 906.0   & --26.2    &
14.5	  & 1.5	    & \nodata &	0.19	  & 0.50      \nl
IRC+10216$^{\rm G}$ & 9.1 & 189.5   & --22.9    &
14.9	  & 1.7	    & \nodata &	0.19      & 0.50      \nl
V Hya	  & 3.2	    & 76.0    & $-16.0^{\rm D}$ &
$30.0^{\rm D}$  & $20^{\rm D}$  & \nodata &	0.10  	  & 0.14      \nl
RT Vir	  & 0.9     & 10.4    & 17.3	  &
8.4       & 1.3	    & \nodata &	0.07 	  & 0.13      \nl
W Hya	  & 2.2	    & 22.6    & 40.6	  &
8.5	  & 1.1	    & \nodata &	0.12	  & 0.25      \nl
W Hya$^{\rm E}$ & 1.2	    & 8.0     & \nodata   &
$7.0^{\rm F}$   & \nodata & \nodata & 0.12      & 0.25      \nl
M57	  & 1.5     & 28.3    &	$-2.0^{\rm F}$  &
$27.0^{\rm F}$  & \nodata & \nodata &	\nodata   & 0.20      \nl
R Cyg	  & 1.3	    & 14.6    & --18.2    &
10.4	  & 1.9	    & \nodata &	0.17  	  & 0.25      \nl
$\chi$ Cyg& 3.0	    & 41.5    &	10.6	  &
9.2	  & 1.4     & 13.5    & 0.19      & 0.35      \nl
V  Cyg    & 2.4	    & 38.0    &	14.2	  &
11.6	  & 1.4     & \nodata & 0.14      & 0.35      \nl
V Cyg$^{\rm G}$ & 0.8     & 11.6    &	\nodata   &
12.0$^{\rm F}$  & \nodata & \nodata &	0.14  	  & 0.35      \nl
CRL 2688  & 10.0    & 288.3   & --34.8    &
23.0	  & 19      & 50.0    &	0.20	  & 0.25      \nl
CRL 2688$^{\rm G}$ & 1.8  & 41.4    &	\nodata   &
22.0$^{\rm F}$  & \nodata & \nodata &	0.20  	  & 0.25      \nl
IRC+40540 & 2.6	    & 46.7    &	-16.8	  &
14.6	  & 1.9     & \nodata & 0.17      & 0.35      \nl
R Cas     & 4.0	    & 76.0    & 24.3	  &
12.3	  & 1.2	    & 21.0    &	0.10      & 0.20 \nl

\tablenotetext{A}{Flux was calculated by integration over whole velocity
range, i.e. with possible wings (see $V_{wing}$).}
\tablenotetext{B}{Those values were obtained by fitting the line shape
given by Eq.1.}
\tablenotetext{C}{The two values of r.m.s. correspond to: $\sigma_1$ -- r.m.s.
of the weighted mean of central position spectrum; $\sigma_2$ -- typical r.m.s.
of the spectra used for constructing the map.}
\tablenotetext{D}{Line shape is not well described by Eq.1 --
fitted values should be treated only as approximations.}
\tablenotetext{E}{SiO (8-7) line.}
\tablenotetext{F}{No line profile was fitted -- line parameters were
estimated by eye.}
\tablenotetext{G}{CS (7-6) line.}
\tablecomments{$T_{max}$ and flux (Columns [2] and [3]) were
taken as highest values in the map
(in some cases telescope was clearly mispointed).
Line profile characteristics in Columns (4)-(6) were obtained using
the central position r.m.s. noise weighted spectra, except M57, where
average of all spectra within  $100"\times100"$ region was used.}
\end{planotable}}

In Table~2 we present the observational properties of the stars'
molecular emission, determined by analysis of the line profile observed at the
central position (with the exception of M57 - see the notes to
Table~2). Because we usually obtained  about 10 central position
observations for each star, the resulting spectra have high signal-to-noise
ratios. As may be seen in the maps we present in
Figures~\ref{fig:rscl}--\ref{fig:rcas}, the central positions do not always
coincide  with the observed maximum of the line intensities,
so we give the flux and line intensity taken from the position at which
the flux is highest.

To obtain the LSR velocity of the star $v_{lsr}$, its expansion velocity
$v_e$ and some measure of the line shape, we fitted (for most of the stars)
the simple line profile proposed by Olofsson et al.~(1993) and given by
\begin{equation}
T_{A}(v)=T_0\left[1-\left(\frac{v-v_{lsr}}{v_e}\right)^2\right]^{\gamma/2},
\label{eq:fit}
\end{equation}
where $v_{lsr},v_e$ are defined above, $T_0$ is the line intensity
at the line center, and $\gamma$ describes the line shape.
The values of $0>\gamma,\; 2\gg\gamma>0,\; \gamma=2$ and $\gamma\gg2$
correspond to a horned, a rectangular, a parabolic and a gaussian-like
line profile (for details see Olofsson et al.~1993 and their Fig.5).
For some stars (indicated in Table~2) the fit was not satisfactory
or no fit at all was attempted.

\begin{figure}[t]
\plotfiddle{fig1.ps}{8cm}{0}{50}{50}{-160}{-90}
\caption{Map of the CO(3-2) emission from OH 231.8+4.2 (continuous line
contours), along with fitted model (dotted line contours).
Crosses indicate observed positions. For details see text.}
\label{fig:fit}
\end{figure}

Some of the line profiles show the presence of wings, sometimes
 strong and very extended (for example OH 231.8+4.2 or VY CMa).
The presence of wings is indicated in the $V_{wing}$ column
in Table~2. Finally, the last two columns give r.m.s. noise of the central
position spectrum and typical r.m.s. deviations of the spectra used for
constructing the map.

In subsequent sections we describe observations of 18 observed
evolved stars, in particular concentrating on the observed spatial
distribution of the molecular emission. The reality of the observed spatial
features in the molecular emission depends on the signal-to-noise ratio
(compare peak intensities in Table~2 with typical r.m.s. deviations of the
spectra used for the map construction) of the maps, but as we
observed stars with rather strong emission, in most of the cases our
maps are clean and we believe the features present at the 10-20\% level
of the total integrated flux are real.

To estimate the real sizes of the molecular envelopes,
for every star except M57 we have fitted to the data a biaxial gaussian,
convolved with the telescope beam pattern. This way we obtained
a FWHM extent of the underlying emission as well as the position angle of the
major axis. For some of the stars, which we indicate in the
description below, the fit was not good, so the values quoted should be
treated only as approximate. The example of the fit to one of the
observed stars, OH231.8+4.2, is shown in Fig.\ref{fig:fit}.

\subsection{R Scl}

\begin{figure}[htb]
\plotfiddle{fig2.ps}{8cm}{0}{47}{47}{-160}{-90}
\caption{Map of CO(3-2) emission from R Scl. Left upper panel shows
map of the total integrated flux, with first contour at 10\% of maximum
flux and next contours also at every 10\%, with 50\% contour marked
with heavy line. Other panels show the maps of flux integrated over
different parts of the line, with velocity ranges showed in brackets. For those
maps first contour is at 5\% of the maximum of total flux, so are the
next contours. Crosses indicate observed positions.
The bottom panel shows the line profile toward the
central star position, along with the best fit to the line shape given
by Eq.1.}
\label{fig:rscl}
\end{figure}

The map of a $1'\times 1'$ region around this C-rich semi-regular variable
(Fig.\ref{fig:rscl}) shows that the CO emission from this star is
well resolved. From the fit to the total integrated flux map of CO J=3-2
emission (Fig.\ref{fig:rscl}, left upper corner) we get an angular extent at
half maximum of $20''\times15''$. The fitted position angle of the major axis
is $75\deg$. The CO J=3-2 line profile towards the star
position may be described as two-peaked, corresponding to partially
resolved, optically-thin emission (but the formal $\gamma$ value from
Eq.\ref{eq:fit} is $0.1$, and there is an enhanced blue horn).

As may be seen in Fig.\ref{fig:rscl}, the observed CO J=3-2
emission clearly departs
from spherical symmetry, both in the form of overall, roughly  east-west
elongation (best seen in the integrated emission in the range
$-30,-23\;km\;s^{-1}$)
and smaller size irregularities. Also, the CO J=3-2 line profile
towards the star position shows significant deviations from
the fitted flat-topped profile, with most noticeable the
blue horn enhancement at $\sim -33\;km\;s^{-1}$. Integrated emission between
--38 and $-30\;km\;s^{-1}$ shows signs of south-north elongation.

\subsection{o Cet}

\begin{figure}[htb]
\plotfiddle{fig3.ps}{8cm}{0}{47}{47}{-160}{-90}
\caption{As in Fig.2 for o Cet (Mira).}
\label{fig:mira}
\end{figure}

Fig.\ref{fig:mira} shows maps of the CO(3-2) emission for this
well-known star. The emission is resolved, with an angular extent
at the half maximum of about $15''\times13''$ and a position
angle of $40\deg$.
The CO(3-2) line profile toward the star is not well described by
Eq.\ref{eq:fit}, so both $\gamma$ and $v_e$ depend strongly
on the part of the line which is fitted.

This star is a subject of the detailed study by Knapp et al.~(1994a),
so here we only indicate the possibility of slight elongation in the
northeast-southwest direction, and also a small north-south shift in the
position of the peak intensity with velocity, showing that the blue-shifted
and red-shifted emission is displaced from the central position
by $\pm4''$. This agrees with the results of the higher resolution
maps of Planesas et al.~(1990a,b).

\subsection{U Cam}

\begin{figure}[htb]
\plotfiddle{fig4.ps}{8cm}{0}{47}{47}{-160}{-90}
\caption{As in Fig.2 for U Cam.}
\label{fig:ucam}
\end{figure}

The total integrated flux map of the CO(3-2) emission of this star
(Fig.\ref{fig:ucam}) has a fitted angular extent at half maximum of
$16''\times13''$, although the fit is not a very good due to the
peculiarities of the emission mentioned below.
The fitted position angle of the major axis is about $20\deg$.
The CO(3-2) line profile toward the star is best fit
by flat-topped parabola ($\gamma=0.4$). As in the case of R Scl, we can see
the blue horn enhancement, but not so strongly pronounced.

The CO(3-2) maps presented in Fig.\ref{fig:ucam} are
quite remarkable. First, there is very clear, roughly south-north
elongation, with an aspect ratio of about 2:1. But there is also a
second elongation visible, roughly perpendicular to the first one, forming
together a T-shaped  envelope. The map of the bluemost part of the
emission line ($-24,-12\;km\;s^{-1}$) shows the presence of secondary
peak, separated by about $25''$ from visible in all maps at the same
position main peak.

\subsection{VY CMa}

VY CMa is the only supergiant star in our sample. Maps of the CO(3-2) emission
appears to be only partially resolved, with the fitted FWHM extent
of $10''$. The line profile towards the star position shows
broad,  strong wings. The formal fit to the central part of the line gives
$\gamma=0.4$.

\begin{figure}[htb]
\plotfiddle{fig5.ps}{8cm}{0}{47}{47}{-160}{-90}
\caption{As in Fig.2 for VY CMa.}
\label{fig:vycma}
\end{figure}

The envelope seems only slightly resolved, although
there is some irregular structure present, clearly visible in the map
of total integrated emission. This is probably due to noise,
as the maps of the emission coming from the center of the line
are much cleaner than maps of the
emission coming from the wings. The position of peak intensities
does not change with velocity and is shifted $10''$ west from
the center of the map; this apparent offset is probably not real and may be
due to bad pointing.

\subsection{OH 231.8+4.2}

Fig.\ref{fig:oh231} shows maps of the CO(3-2) emission, which appears
to be resolved. From the total integrated flux map of CO(3-2)
emission we get an angular extent at half maximum of $15''\times1''$,
i.e. the emission is unresolved in the direction perpendicular to the
major axis. The position angle of the major axis is about $5\deg$.
The fit to the emission is shown in Fig.\ref{fig:fit}.
The CO(3-2) line profile towards the star position is
remarkable, with wings extending over $100\;km\;s^{-1}$. The formal fit
of the line shape given by Eq.\ref{eq:fit} depends strongly on which
velocity range is used for the fit, so the formal $\gamma=11$ only indicates
the gaussian-like shape of the line, i.e. corresponds to unresolved,
optically thick emission.

\begin{figure}[htb]
\plotfiddle{fig6.ps}{8cm}{0}{47}{47}{-160}{-90}
\caption{As in Fig.2 for OH 231.8+4.2.}
\label{fig:oh231}
\end{figure}

Maps of different parts of the line show clear bipolar structure --
the peak intensity moves from the north to the south with velocity.
The maximal displacement from the central position is about $\pm8''$.
The elongation clearly visible in the map of the total integrated flux
seems be mainly due to this shift, but individual maps still show some
north-south elongation,
especially well seen in the maps of the wings . Existence of the two
components in the emission, one unresolved with only moderate outflow
velocities and another showing clear bipolar structure and outflow velocities
greater then $100\;km\;s^{-1}$ agrees with  the CO(1-0) observations by
Morris et al.~(1987).

\subsection{RS Cnc}

The total integrated flux of the CO(3-2) emission of this star
(Fig.\ref{fig:rscnc}) has the fitted  angular extent at half maximum
of $12''\times8''$. The position angle of the major axis is $70\deg$.
The CO(3-2) line profile toward the star is not well described
by Eq.\ref{eq:fit}, but formal fit to the central part
of the line gives $\gamma=4.6$, i.e. the line is gaussian-like (unresolved
and optically thick). This line shape is similar to that of o Cet.

\begin{figure}[htb]
\plotfiddle{fig7.ps}{8cm}{0}{47}{47}{-160}{-90}
\caption{As in Fig.2 for RS Cnc.}
\label{fig:rscnc}
\end{figure}

As may be seen in Fig.\ref{fig:rscnc}, the CO(3-2) maps
indicate a possible slight west-east elongation, which is confirmed
by the fitting procedure described earlier. There is no indication
of changes in the position of peak intensity with velocity.

\subsection{R Leo}

This nearby Mira variable was mapped at $9''$ spacing and a $7\times7$ grid.
The maps of the molecular emission are presented in Fig.\ref{fig:rleo}.
{}From the map of the total integrated flux in the CO(3-2) line we get the
fitted angular extent of the emission at half maximum of  $14''\times11''$
and the position angle of the major axis $125\deg$.
The spectrum towards the star position reveals in addition to the CO(3-2) line,
a weaker SiO(8-7) line detected in the image sideband, apparently
lying just bluewards of CO(3-2) line. As R Leo was mapped twice,
in January 1990 ($5\times5$ map) and May 1990 ($7\times7$ map), we present
here only bigger maps. We used the spectrum from January 1990 to fit a
theoretical line profile as this spectrum was the result of averaging
about 20 individual spectra, much more that were taken in May 1990. The best
fit was achieved with $\gamma=2.1$, i.e. inverted parabola fit,
but there is clearly visible red horn enhancement at about $3\;km\;s^{-1}$
present in spectra taken in both epochs.

\begin{figure}[htb]
\plotfiddle{fig8.ps}{8cm}{0}{47}{47}{-160}{-90}
\caption{As in Fig.2 for R Leo. The lower panel contains
line profiles of CO(3-2) line and SiO(8-7) line (left).
Map of the emission in the SiO(8-7) line is presented in the middle
panel on the right.}
\label{fig:rleo}
\end{figure}

As we mentioned above, molecular emission from R Leo was mapped twice,
once on $5\times5$ grid and second time on $7\times7$ grid. We found
good agreement between the maps taken in both epochs. First, there is
clearly present southeast-northwest elongation, visible also in the map
of SiO(8-7) emission. Second, there is systematic shift in the position
of the peak intensities with the velocity, from west for approaching
parts of the envelope to east for receding emission, what indicates
a bipolar character of the emission. It is interesting to notice that
the peak intensity of the SiO(8-7) emission is shifted even more to the east
than the redmost part of the CO(3-2) emission.

R Leo was also mapped by Bujarrabal \& Alcolea (1991, hereafter: BA),
but it is rather difficult to compare our results with their smaller,
non-uniformly sampled map. We confirm the presence of the two horn structure
seen by them in CO(2-1) line profile in our data taken in May 1990.

\newpage

\subsection{IRC +10216}

Maps of this well-known carbon-rich star are presented in
Fig.\ref{fig:irc10}. In this case we were able to map
along with the CO(3-2) emission also emission in the CS(7-6)
line. The total integrated flux of CO(3-2) emission is well resolved and
has the fitted angular dimension at half maximum of $24''\times16''$.
The position angle of the major axis is about $100\deg$.
Line profiles toward the star position of both CO(3-2) and CS(7-6)
are best fitted with $\gamma\approx1.6$.

\begin{figure}[htb]
\plotfiddle{fig9.ps}{8cm}{0}{47}{47}{-160}{-90}
\caption{As in Fig.2 for IRC +10216. The lower panel contains
line profiles of CO(3-2) line (left) and CS(7-6) line (right).
Map of the emission in the CS(7-6) line is presented in the middle
panel on the right.}
\label{fig:irc10}
\end{figure}

As this star is a subject of detailed study elsewhere (Knapp et al.~1994b),
here we want only to point out the clearly visible west-east elongation,
present in both lines. The position of peak intensity does not change
with velocity. There is also marked asymmetry in the CO(3-2) emission,
visible at the level of 20\% of maximum flux, namely enhanced
emission south of the star position (best visible between --33 and --19
$km\;s^{-1}$).

This star was previously mapped by Troung-Bach, Morris \& Nguyen-Q-Rieu
(1991), where from maps of the  CO(1-0) and CO(2-1) emission  they conclude
that the observations indicate a spherical symmetry for the envelope.

\subsection{V Hya}

Maps of the CO(3-2) emission from this star, obtained using a $9\times9$
grid with $9''$ spacing, are presented in Fig.\ref{fig:vhya}. The fitted
angular extent of the emission at the half maximum of the total integrated flux
is $15''\times12''$. The position angle of the major axis is about $175\deg$.
The CO(3-2) line profile towards the star position is
best fitted with high value of $\gamma\approx20$., i.e. gaussian-like line
shape, although there is present strong red horn enhancement at about
$11\;km\;s^{-1}$.

The maps reveal a number of interesting features. The map of the
total integrated flux shows very extended emission on the 10\% level, unlike
the case of other stars we observed when the separations between contours
were rather similar. The same map, as well as individual position-velocity
maps, shows north-south elongation, confirmed by our fitting
procedure. Finally, there is a clearly visible
trend in the position of the peak intensities as a function of velocity,
with emission from the blue parts of the line being shifted to the west
and emission from the red parts of the line being shifted to the east.
The amplitude of this effect, indicating a bipolarity in the outflow,
is about $3''$.

\begin{figure}[htb]
\plotfiddle{fig10.ps}{8cm}{0}{47}{47}{-160}{-90}
\caption{As in Fig.2 for V Hya}
\label{fig:vhya}
\end{figure}

This star was previously mapped by Tsuji et al.~(1988) and
Kahane, Maizels \& Jura~(1988) in CO(1-0) line and both groups concluded
that the emission has a bipolar nature, as confirmed here
with better sampled and higher resolution maps of CO(3-2) emission.

\subsection{RT Vir}

Fig.\ref{fig:rtvir} shows $5\times5$ maps, made with the $9''$ spacing,
of the CO(3-2) emission, which seems to be partially resolved.
The total integrated flux map has a fitted angular
extent at  half maximum of $15''\times8''$. The fitted position angle of the
major axis is about $30\deg$. CO(3-2) line profile towards the star
position is best fitted by slightly flat-topped parabola ($\gamma=1.3$).

\begin{figure}[htb]
\plotfiddle{fig11.ps}{8cm}{0}{47}{47}{-160}{-90}
\caption{As in Fig.2 for RT Vir.}
\label{fig:rtvir}
\end{figure}

Maps of different parts of the line seems to indicate that the emission
is elongated, roughly northeast to southwest. Positions of the peak
intensities seems to move slightly with velocity by about $3''$, but in rather
incoherent fashion.

\subsection{W Hya}

$5\times5$ maps of the molecular emission from this O-rich star are presented
in Fig.\ref{fig:whya}. From the map of the total integrated flux
in the CO(3-2) line we get the fitted angular diameter at half maximum of
$13''\times11''$ at a position angle of $130\deg$. The CO(3-2) line is best
fit with $\gamma=1.1$, and strong emission in the SiO(8-7) line at 343 GHz
is also seen. The CO(3-2) line shows the presence
of strong blue horn enhancement, similar to that of U Cam, with a rather
weak red horn also present.

\begin{figure}[htb]
\plotfiddle{fig12.ps}{8cm}{0}{47}{47}{-160}{-90}
\caption{As in Fig.2 for W Hya. The lower panel contains
line profiles of CO(3-2) line and SiO(8-7) line (left).
Map of the emission in the SiO(8-7) line is presented in the middle
panel on the right.}
\label{fig:whya}
\end{figure}

The maps of molecular emission shown in Fig.\ref{fig:whya} reveal the
presence of two components. The blue part of the CO(3-2)
($32,41\;km\;s^{-1}$) and SiO(8-7) lines are elongated from
northeast to southwest, while the red part of CO(3-2) line emission
($41,51\;km\;s^{-1}$) is elongated in roughly perpendicular direction,
from southeast to northwest. The peak intensity of the redmost part of the
CO(3-2) line is shifted by about $5''$ east from otherwise steady peak
intensities in other velocity ranges and in the SiO(8-7) line.

\subsection{M57}

Maps of the molecular emission from the well-known Ring Nebula (M57)
are presented on Fig.\ref{fig:m57}. Because of the complicated structure, both
grey scale and contour maps of the integrated emission are presented.
The emission forms an elliptical ring slightly offset from the center position
and with angular dimension at half maximum of $100''\times60''$.
The position angle of the major axis is about $60\deg$. No fit was
attempted to the line profiles, which often show multiple components.
We show the spectrum corresponding to the position of
maximum total integrated flux at (--40,--10),
and also the average of all the spectra within $\pm50''$ of the center.

\begin{figure}[htb]
\plotfiddle{fig13.ps}{8cm}{0}{47}{47}{-160}{-90}
\caption{As in Fig.2 for M 57. The map of the total integrated
flux was in addition presented using greyscale shading
(starting at 20\% of the maximum) as to clarify complicated
spatial structure of this planetary nebula visible in CO emission.
The lower panel contains two spectra: spectrum obtained at the offset
position (-40,-10), which has the maximum integrated flux (left) and
average of all the spectra within the region $\pm50''$ from the center
(right). For M57 no fit to spectral lines was attempted.}
\label{fig:m57}
\end{figure}

Although the emission from M57 is weak, the maps show a number
of remarkable features.
First, there is a clearly visible depletion of CO around
the central position. Further, the emission shows systematic
changes with velocity. The bluemost part of the emission
shows concentration towards the central position and comes
from the western part of the nebula. Emission from the line center
$[-8,2]\;km\;s^{-1}$ comes from outer parts of the nebula.
This velocity range shows the presence of two ``hot spots'',
one around (--45,--15) and other one at (+35,+25). Emission coming
from the receding part of the nebula is concentrated towards the
central position and is visible mostly from the eastern part of the nebula.
All maps show a high degree of irregularity.

M57 was observed by Bachiller et al.~(1989) in the CO(2-1)
and CO(1-0) lines. Our results agree very well  with their
observations of CO(2-1) emission at $14''$ resolution.
Following their suggestions that the observed velocity and spatial
structure indicates that the emission is coming from a hollow cylinder (
barrel), inclined with respect to the line of sight, we tried to
construct a kinematical model of the nebula. We found
that we  can reach some agreement when modelling the emission as
a thin-walled, elongated cylinder, but we were not able
reproduce the observed hot spots within the framework of uniform density
and velocity kinematical models.

\subsection{R Cyg}

A map of a $36''\times36''$ region around this Mira variable is presented
in Fig.\ref{fig:rcyg}. The $5\times5$ sampling grid had a pixel spacing
of $9''$. The emission seems to be slightly resolved, with the fitted angular
extent at the half maximum of total integrated flux of about $14''\times9''$
and the position angle of $135\deg$, although in this case fit by biaxial
gaussian is very approximate.
The CO(3-2) line profile towards the central position is best fitted with
inverted parabola ($\gamma=1.9$), although the line shows considerable
irregularities.

\begin{figure}[htb]
\plotfiddle{fig14.ps}{8cm}{0}{47}{47}{-160}{-90}
\caption{As in Fig.2 for R Cyg.}
\label{fig:rcyg}
\end{figure}

\newpage

\subsection{$\chi$  Cyg}

Maps of CO(3-2) emission from this star, obtained using a $7\times7$ grid
with $10''$ spacing, are presented in  Fig.\ref{fig:chicyg}.
The fitted angular extent at the half maximum of the total integrated flux
is $16''\times10''$. The fitted position angle of the major axis is  $115\deg$.
The CO(3-2) line profile towards the star position is best fitted with
$\gamma=1.4$. There is a weak wing present at the red side of the line.

\begin{figure}[htb]
\plotfiddle{fig15.ps}{8cm}{0}{47}{47}{-160}{-90}
\caption{As in Fig.2 for $\chi$ Cyg.}
\label{fig:chicyg}
\end{figure}

The maps show a number of interesting features.
First, the emission is clearly extended and elongated
from southeast to northwest. Second, there is striking structure
present in some velocity-position maps, pointing towards north-west.
Finally, peak intensities show bipolarity, with emission shifting
towards the west with velocity. The amplitude of this effect is about $4''$.

$\chi$ Cyg was previously mapped by BA in the CO(2-1) line,
but they do not find any significant deviations from spherical symmetry.

\newpage

\subsection{V Cyg}

We mapped V Cyg using a $7\times7$ sampling grid with $10''$ spacing,
and the resulting maps of molecular emission are presented in
Fig.\ref{fig:vcyg}. From the map of total integrated CO(3-2)
emission we get the fitted angular extent at half maximum of $15''\times9''$.
The position angle of the major axis is $25\deg$. The
spectrum towards the star's position shows  CS(7-6) line.
The CO(3-2) line profile was best fit with $\gamma=1.4$. There is a weak
wing present at the red part of the line, but we do not integrate
emission with $v<5\;km\;s^{-1}$ as there is a narrow emission line at about
$-1\;km\;s^{-1}$, probably Galactic in origin,  which is strongest
to the east from the star.

\begin{figure}[htb]
\plotfiddle{fig16.ps}{8cm}{0}{47}{47}{-160}{-90}
\caption{As in Fig.2 for V Cyg. The lower panel contains
line profiles of CO(3-2) line and CS(7-6) line (right).
Map of the emission in the CS(7-6) line is presented in the middle
panel on the right.}
\label{fig:vcyg}
\end{figure}

The maps of molecular emission are fairly noisy,
maps, but there are a number of features we want to comment on.
CO(3-2) emission is clearly resolved and shows elongation from
northeast to southwest. There is an analogous trend in the position
of peak intensities, which move with velocity by about $\pm3''$,
possibly indicating bipolarity of the outflow. CS(7-6) emission
is weaker and therefore noisier, but we believe the east-west elongation
seen in the middle-right panel is real. Our observations agree with the
results of BA (their Fig.10).

\subsection{CRL 2688}

Maps of the molecular emission from this carbon-rich star are presented
in Fig.\ref{fig:crl2688}. From the map of the total integrated CO(3-2) flux
we get the fitted diameter of the half maximum emission to be about $12''$.
The spectrum taken towards the star's position shows CO(3-2) line
and also CS(7-6) emission.  The line profile of CO(3-2) emission,
discussed by Young et al.~(1992), was in the central part
best fitted by a gaussian-like shape (Table~2). No fit was attempted
to the weaker and noisier CS(7-6) line.

\begin{figure}[htb]
\plotfiddle{fig17.ps}{8cm}{0}{47}{47}{-160}{-90}
\caption{As in Fig.2 for CRL 2688. The lower panel contains
line profiles of CO(3-2) line and CS(7-6) line (left).
Map of the emission in the CS(7-6) line is presented in the middle
panel on the right.}
\label{fig:crl2688}
\end{figure}

The maps show that the emission is basically spherically symmetric,
with suggestion of extension toward the north-east.
There is possible bipolarity in the positions of the peak
intensities, with emission shifting from the east for red parts of the line
to the west for blue parts of the line, with an amplitude of about $2''$.

CRL2688 was mapped several times before in different wavelengths and
the results were rather inconsistent. The spherical symmetry we observed
agrees with the results of observations by Truong-Bach et al.~(1990),
with somewhat better resolution in the CO(2-1) line, and with the
results of Yamamura et al.~(1994) obtained with Nobeyama Millimeter
Array.

\subsection{IRC +40540}

Maps of the CO(3-2) emission from this star are presented in
Fig.\ref{fig:irc40}. The total integrated emission seems to
be partially resolved and has the fitted angular extent at half maximum
of $12''\times6''$. The position angle of the major axis is about $60\deg$.
The line profile is best fitted with inverted parabola
($\gamma=1.9$). There is an enhancement in the line emission
at about $18\;km\;s^{-1}$.

\begin{figure}[htb]
\plotfiddle{fig18.ps}{8cm}{0}{47}{47}{-160}{-90}
\caption{As in Fig.2 for IRC +40540.}
\label{fig:irc40}
\end{figure}

Maps of the CO(3-2) emission  clearly show that the emission
is elongated in the northeast-southwest direction. Individual
velocity-position maps show the presence of jet-like structures,
the clearest one of which, pointing towards northeast and seen between
$-24$ and $-14\;km\;s^{-1}$ may be associated with mentioned above
enhancement in the line profile. The peak intensities change their
position, but in rather incoherent fashion.

\subsection{R Cas}

Maps of the CO(3-2) emission from R Cas, made using a $5\times5$ grid with
$9''$ spacing, are presented in Fig.\ref{fig:rcas}. The half maximum
contour of the total integrated flux is elongated and has fitted
angular extent of $13''\times10''$. The fitted position angle of the major axis
is $70\deg$. The line profile
towards the star position is best fitted with $\gamma=1.2$, with some
red-wing enhancement and weak wind present in the blue part of the line.

\begin{figure}[htb]
\plotfiddle{fig19.ps}{8cm}{0}{47}{47}{-160}{-90}
\caption{As in Fig.2 for R Cas.}
\label{fig:rcas}
\end{figure}

The maps presented indicate east-west elongation, visible in all individual
velocity ranges. The peak intensity of the bluemost part of the CO(3-2) line
($4,15\;km\;s^{-1}$) is shifted by about $5''$ from otherwise steady peak
intensities in other velocity ranges, and also this part of the line
shows the clearest elongation.

Molecular emission from this star was previously mapped by
BA, but they do not indicate any deviations from
spherical symmetry, even if their CO(2-1) map (their Fig.3) seems
to suggest some elongation, consistent with what we observed.
The multiple horn structure, which they observe in the CO(2-1) line profile,
seems to be present also in our observations of the CO(3-2) line, seen
in Fig.\ref{fig:rcas}.

\subsection{Discussion}

The observations described above show that the majority of the observed
stellar envelopes show deviations from circular symmetry when mapped
in molecular line emission. We are reasonably confident that these deviations
are real; for example, the derived position angles are randomly distributed.
In some cases, our results agree with these of previous observations,
but in others they do not. The CO(3-2) line traces higher density gas and
is possibly more sensitive to the envelope structure.

\section{Envelope Radii}

CO in molecular outflows is destroyed in the outer parts of stellar
envelopes by photodestruction by interstellar ultraviolet radiation,
and this process limits the effective sizes of the envelopes as
measured by the CO line (Mamon, Glassgold and Huggins 1988).  The CO
extent of the envelope is given approximately by
\begin{equation}
R_{\rm CO} = 5.4 \times 10^{16} \left({{\mathaccent 95 M}
\over {10^{-6}}}\right)^{0.65} \left({{V_e} \over {15}}\right)^{-0.55}
\left({{f} \over{8 \times 10^{-4}}}\right)^{0.55} + 7.5 \times 10^{15}
\left({{V_e} \over {15}} \right)   \;cm,
\end{equation}
where the mass loss rate $ \mathaccent 95 M$ is in $ M_{\odot}
 \; yr^{-1}$, $ V_e$ is in $ km\;s^{-1}$ and $f$, the fractional
abundance of CO at the inner edge of the wind is $f = \rm n(CO)/n(H_2)$.
This interpolation formula is described by Knapp et al.~(1994c); the
first term is the self-shielding term for CO, and the second is the mean
free-flight distance of an unshielded CO molecule before it is destroyed.

The mass loss rates for each of the stars (except M57) were first calculated
from the data in Table 2 using the models described by Knapp et al.~(1994c),
based on earlier work by Knapp \& Morris~(1985).
Table 3 lists for each star the assumed distance in $pc$ (taken from
Knapp et al.~1994c) and the mass loss rate. The  relative CO abundances
were assumed to be $f = 5 \times 10^{-4}$ for
oxygen stars, $ 6.5 \times 10^{-4}$ for S stars and $ 10^{-3}$ for
carbon stars.  The effective photodissociation radius was then calculated
from this mass loss rate and the wind velocity from Table 2, and is
listed in Column 4 of Table 3.

The observed half-power diameters of the stars are listed in the next column.
They were calculated using the geometric means of the angular diameters given
in the previous section.  An upper limit of $ 10''$ was assumed for any
unresolved source.  The line emission from circumstellar envelopes is
brightest in the center and we have approximated the brightness with a
gaussian distribution; we might therefore expect the total source
diameter to be about twice as large as the half power diameter, in
which case the linear half power diameter should approximately
equal the source radius.

\begin{planotable}{lrlrrrl}
\tablewidth{40pc}
\tablecaption{Mass Loss Rates and Envelope Sizes}
\tablehead{
\colhead{Star}        & \colhead{Distance} &
\colhead{$\dot{M}$}	& \colhead{$R_{\rm CO}$ (mod.)}  &
\colhead{$R_{\rm CO}$ (obs.)} & \colhead{$R_{\rm dust}$} &
\colhead{$M_{\rm env}$} \vspace{0.05cm} \\
\colhead{} 	        & \colhead{[$\,pc\,$]} &
\colhead{[$\,M_{\odot}\;yr^{-1}\,$]} & \colhead{[$\,10^{16}\;cm\,$]} &
\colhead{[$\,10^{16}\;cm\,$]} & \colhead{[$\,10^{17}\;cm\,$]} &
\colhead{[$\,M_{\odot}\,$]} }
\startdata
R Scl	   & 330  & $1.8\times10^{-6}$ & 9.6  & 8.6 & 19.0	&
$7.1\times10^{-2}$ \nl
o Cet	   & 80   & $2.6\times10^{-7}$ & 3.0  &	1.7 & 1.6	&
$1.7\times10^{-3}$ \nl
U Cam	   & 430  & $3.5\times10^{-6}$ & 12.0 &	9.3 & \nodata	&
\nodata \nl
VY CMa	   & 1500 & $3.5\times10^{-4}$ & 122.0 & 22.0 & \nodata &
\nodata  \nl
OH 231.8+4.2 & 1300 & $3.0\times10^{-4}$ & 79.0 &	$\leq$30.0 & \nodata &
\nodata \nl
RS Cnc	   & 70   & $6.0\times10^{-8}$ & 1.5  &	1.0 & 3.6	&
$1.4\times10^{-3}$ \nl
R Leo  	   & 120  & $1.8\times10^{-7}$ & 2.3  &	2.2 & 4.1	&
$3.0\times10^{-3}$ \nl
IRC+10216  & 150  & $2.0\times10^{-5}$ & 44.0 &	4.4 & \nodata	&
\nodata \nl
V Hya	   & 380  & $5.0\times10^{-6}$ & 13.0 &	8.0 & \nodata	&
\nodata \nl
RT Vir	   & 110  & $8.0\times10^{-8}$ & 1.5  &	1.8 & 4.1	&
$1.2\times10^{-3}$ \nl
W Hya	   & 90   & $8.0\times10^{-8}$ & 1.5  &	1.6 & 8.5	&
$2.5\times10^{-3}$ \nl
R Cyg	   & 900  & $3.5\times10^{-6}$ & 14.0 &	15.0 & 39.0	&
$4.1\times10^{-1}$ \nl
$\chi$ Cyg & 140  & $2.8\times10^{-7}$ & 3.2  &	2.6 & \nodata	&
\nodata \nl
V Cyg	   & 330  & $9.5\times10^{-7}$ & 7.4  &	5.7 & \nodata	&
\nodata \nl
CRL 2688   & 1000 & $2.0\times10^{-4}$ & 97.0 &	18.0 & \nodata 	&
\nodata \nl
IRC+40540  & 600  & $4.5\times10^{-6}$ & 17.0 &	7.6 & \nodata	&
\nodata \nl
R Cas      & 160  & $1.0\times10^{-6}$ & 5.3  &	2.7 & 6.2	&
$1.6\times10^{-2}$
\end{planotable}

As can be seen from Table 3, the agreement between the predicted and observed
source radii is generally quite good.  The worst discrepancies come from
the stars with the largest mass loss rates, such as IRC+10216, CRL 2688
and VY CMa. This observation could indicate that mass loss has not been going
on for very long for these stars, but it is more likely that the observations
are not sensitive to the weak emission from cold gas at large distances
from the star.

The last two columns of Table 3 list the observed angular radius of the dust
emission measured from the IRAS data by Young et al. (1993) and the total
envelope mass calculated (assuming that the mass loss rate has been
constant) from this radius, the outflow speed $ V_e$ in Table 2 and the
value of the mass loss rate from column 3.  It can be seen that, indeed,
in all cases the observed CO envelope radius is much smaller than that
given by the dust emission, confirming the model of Mamon, Glassgold and
Huggins (1988).

\section{Conclusions}

We find that the majority of the observed by us 18 evolved stars
show clear deviations from spherical symmetry, in the form of elongated
emission, bipolar emission and smaller scale irregularities. This may
lead  to the conclusion that a substantial fraction of all stars in the AGB
phase may lose matter in a non-spherical manner. One needs to look at other
mechanisms for distorting the envelope than the action of a binary companion.
A promising candidate is convection -- there may be only a few convection
cells in these enormous stars.  One would expect that when the optical
photospheres of these stars are resolved the stars themselves will not be
spherically symmetric.

Some of the stars we observed are strong enough to be observed
in higher than CO(3-2) transitions, which would increase the spatial
resolution with which one could map the molecular emission with. Mira (o Ceti)
is a strong source, only partially resolved in our data. IRC+10216
is another good candidate. Observations with millimeter wave interferometers
would also yield better maps even at lower $J$\/ transitions. We expect higher
resolution maps to reveal the asymmetries we see more clearly.

\acknowledgments{We would like  to thank R.~H.~Lupton for his valuable help
with the SM plotting package. We are also indebted to the staff of the CSO
for their help and support during the observing runs over which these data
were collected. Research at the CSO is supported by NSF grant
AST90-15755. This work was also partially supported by NSF grant
AST90-14689 to Princeton University. This research has made use of the
SIMBAD database, operated at CDS, Strasbourg, France.}

\end{document}